\documentclass[10pt, final, conference, letterpaper, twocolumn, twoside]{IEEEtran}
\usepackage{latex8}
\usepackage{times}
\usepackage{subfigure,algorithm}
\usepackage{amsmath}

\usepackage{amsfonts}

\usepackage[dvips]{graphicx}
\usepackage{fancybox}

\usepackage{color}
\usepackage{eso-pic}

\graphicspath{{./Figures/}}

\newcommand{\ben}{\begin{enumerate}}
\newcommand{\een}{\end{enumerate}}
\newcommand{\be}{\begin{equation}}
\newcommand{\ee}{\end{equation}}
\newcommand{\bea}{\begin{eqnarray}}
\newcommand{\eea}{\end{eqnarray}}
\newcommand{\bc}{\begin{cases}}
\newcommand{\ec}{\end{cases}}
\newcommand{\bi}{\begin{itemize}}
\newcommand{\ei}{\end{itemize}}

\renewcommand{\and}{\hspace{0.4cm}}

\begin{document}
\AddToShipoutPicture*{\small
\sffamily\raisebox{1.8cm}{\hspace{1.8cm}1-4244-2575-4/08/\$20.00
\copyright2008 IEEE}}

\title{Informed Network Coding for Minimum Decoding Delay}

\author{Rui A. Costa$^1$ \and Daniele Munaretto$^2$ \and Joerg Widmer$^2$ \and Jo\~ao Barros$^1$\\[0.2cm]
$^1$Instituto de Telecomunica\c{c}\~oes,
Faculdade de Ci\^encias da Universidade do Porto, Portugal\\\{ruicosta, barros\}@dcc.fc.up.pt\\
\and
\\
$^2$DoCoMo Euro-Labs, Munich, Germany \\lastname@docomolab-euro.com\\
}

\maketitle
\thispagestyle{empty}

\begin{abstract}
Network coding is a highly efficient data dissemination mechanism for
wireless networks. Since network coded information can only be
recovered after delivering a sufficient number of coded packets, the
resulting decoding delay can become problematic for delay-sensitive
applications such as real-time media streaming. Motivated by this
observation, we consider several algorithms that minimize the decoding
delay and analyze their performance by means of simulation. The
algorithms differ both in the required information about the state of
the neighbors' buffers and in the way this knowledge is used to decide
which packets to combine through coding operations. Our results show
that a greedy algorithm, whose encodings maximize the number of nodes
at which a coded packet is immediately decodable significantly
outperforms existing network coding protocols.

\end{abstract}

\section{Introduction}

The basic idea of network coding \cite{inftheory,practical_coding}, by
which nodes transmit packets that result from joint encoding of
multiple original information units, has led to communication
protocols that are applicable in a wide range of wireless
communication scenarios \cite{Fragouli2006Primer}. The gains brought
by network coding are most evident in applications involving multicast
or broadcast sessions (in which messages are intended for multiple
destination nodes) in combination with physical layer broadcast (in
which neighboring nodes can overhear potentially useful information).

For delay-sensitive applications such as media streaming, it is not
desirable for receivers to have to wait for the arrival of several
packets until they are able to decode the sent data. Instead, we would
like for a packet to be immediately decodable given only the
information units already available at those nodes. Moreover, each
packet should be useful for as many neighbors as possible, thus
minimizing the required number of transmissions. Similar
considerations hold for distributed systems with highly constrained
nodes, such as sensor networks, where it may well be impossible to
store a large number of coded packets and to decode them using
Gaussian elimination.

In general, there is a tradeoff between delay, throughput and
end-to-end quality, and different codes can try to optimize either one
of these performance metrics. Priority Encoded
Transmission~\cite{albanese1996pet} provides graceful degradation by
specifying different levels of coding (and consequently different
minimum numbers of packets required for decoding) depending on the
content and the priority of the underlying information units. Fountain
codes (e.g.~Raptor codes~\cite{shokrollahi2006rc}) offer very low
coding overhead and are (asymptotically) rate optimal when
transmitting over erasure channels.  However, decoding $n$ packets is
only possible \emph{after} $n+\epsilon$ coded packets have been
received. A fraction of the encoded symbols can be decoded earlier
albeit at the cost of significant overhead~\cite{sanghavi2007ipr}.

We focus our attention on algorithms that allow \emph{immediate}
decoding of each incoming packet.  This stringent delay requirement
comes at the cost of a reduction in throughput in the sense that
broadcasting coded packets brings new information to fewer neighbors
than fountain coding or priority encoded transmission would allow.
Fostering early or immediate decoding requires an algorithm that
decides which and how many information units or symbols\footnote{These
  two expressions shall be used interchangeably throughout the paper.}
should be combined in each new packet that is to be transmitted.
 
Adequate design of such an algorithm is highly dependent on
the state information available at the nodes. Complete lack of state
information is likely to occur in highly dynamic networks, such as
mobile sensor networks, where the overhead of tracking a changing
neighborhood would be prohibitive. In case a node has no information
about the packets that have already been recovered by its neighbors,
the algorithm can only optimize how many information units to combine
(i.e., the codeword degree~\cite{GC}).  Each node simply combines
randomly chosen information received from other nodes with its own
information units, until the desired codeword degree is reached. The
algorithm needs to find the right tradeoff between a high codeword
degree that ensures that coded packets bring new information to many
of the neighbors, and a low codeword degree that allows packets to be
decoded immediately using only the information that is locally
available. An analysis of optimum degree distributions with respect to
network dynamics and topology was carried out
in~\cite{Resilient_Coding}.

When information about the data recovery status of neighboring nodes
is available, it is possible to employ more sophisticated coding
algorithms. One such instance is presented in \cite{nc_xor}, which
proposes a protocol for unicast routing in wireless mesh
networks. Routers combine packets opportunistically from different
sources in order to increase the diversity of the information content
of each transmission. A node chooses the symbols to combine based on
the content of the neighbors' buffers. This form of state information
is piggybacked onto data packets and also extrapolated from past loss
rate measurements and overheard packets. The procedure ensures that
coded packets are immediately decodable at the next hop with very high
probability.

Although the protocol in \cite{nc_xor} targets unicast traffic, very
similar considerations also apply to broadcast.
\cite{Online_broadcasting_with_NC} analyzes a number of simple
heuristics for the online and offline version of the problem.

Intrigued by the behavior of network coding protocols for one-to-all
or all-to-all information dissemination, we compare the performance of
several existing coding algorithms with various levels of buffer state
information on neighboring nodes. We further propose two new
protocols, one that maximizes the number of immediately decodable
packets in a greedy fashion, and one that attempts to equalize the
number of recovered information units among neighbors. The proposed
schemes operate under the assumption that neighboring nodes exchange
information about which symbols they have already recovered, for
example by appending this information every time they send a data
packet.

We show that the proposed schemes outperform existing algorithms in
various scenarios of interest.  We first consider a simple decoder
that discards all incoming packets that cannot be decoded
immediately. We then proceed with a characterization of the
performance gains induced by a more complex decoder that buffers all
received packets and uses Gauss-Jordan elimination to recover the sent
data.

The remainder of the paper is structured as follows. In
Section~\ref{sec:related} we discuss related coding algorithms that
shall serve as a baseline for the subsequent
comparison. Section~\ref{sec:algorithms} describes our two network
coding algorithms, whose performance is highlighted in
Section~\ref{sec:results}, where simulation results are used to
compare their performance to that of existing
solutions. Section~\ref{sec:conclusions} concludes the paper.

\section {Review of Existing Coding Algorithms}
\label{sec:related}

Network coding enables nodes to transmit packets that are a
combination of multiple original data packets. For this purpose,
coding operations are applied to symbols (or sequences of bits), which
can be viewed as elements of a finite field.  When combining a set of
packets, the same operations are applied to all of the symbols that
form these packets and consequently the output packet has the same
length as the input packets. Linear codes have been shown to be
sufficient to achieve the multicast capacity of a network and can be
easily implemented in practice.

Network coding schemes often generate each coded packet in a
randomized fashion by taking into consideration all the packets that
are available in the send buffer.  This approach requires no knowledge
about the recovery status of the neighboring nodes and no
pre-established degree distribution (as explained later).  The most
prominent representative of this class of schemes is the {\it Random
  Linear Network Coding} (RLNC) algorithm presented in \cite{hoko:03},
where the coefficients used to generate the output linear combination
are chosen randomly from the prescribed field.

The algorithms presented in the next section are based on the simplest
form of network coding, whose single operation is binary addition (in
contrast with RLNC, which generally requires addition and
multiplication in higher fields). Since the symbols are elements of
the binary field, adding two packets amounts to the bit-wise XOR of
their symbols. It is worth mentioning, however, that the design
considerations presented in this paper also hold for fields of larger
size.

Regarding the decoding process, we consider two different mechanisms:
(1) a very simple decoding scheme, which uses only immediately
recovered symbols for decoding a new symbol from a received packet;
and (2) the full decoding scheme, which performs Gaussian-Jordan
elimination based on both coded and undecoded packets that are stored
in a node's buffer.

As an example of a RLNC scheme, we now give a formal definition of the algorithm
presented in \cite{Online_broadcasting_with_NC}. This Systematic Random
Network Coding scheme was designed for the scenario of a
source broadcasting to $n$ receivers over independent erasure
channels.\\

\textbf{Systematic Random Network Coding}
\cite{Online_broadcasting_with_NC}: \emph{The algorithm starts by
  sending every packet once in uncoded form. After this first phase,
  the algorithm computes the output packet as a random linear
  combination of all the (uncoded) packets in the buffer.}\\

Since the systematic algorithm produces random linear combinations of
all the packets in the send buffer, the original symbols can only be
recovered by means of Gauss-Jordan Elimination and only after enough
independent linear combinations have been gathered by the destination
node.  In general, these coded symbols cannot be recovered at an
earlier stage thus defeating a simple decoder, which discards all
packets that are not immediately decodable using solely the already
decoded information. It is therefore necessary to limit the number of
symbols combined at each step, also called the \emph{codeword degree}.

\subsection{Degree distribution based algorithm}

Network coding schemes that combine all packets in the buffer are
not always optimal in terms of performance, as observed in \cite{GC}.
The authors show that depending on the number of recovered packets $r$
at a specific node, there exists an optimal number of packets to
combine to maximize the number of decodable packets at each instant in
time. More precisely, defining the {\it codeword degree} as the number
of original symbols which are jointly encoded to form a coded
packet, the authors of \cite{GC} determine the optimum \emph{degree
  distribution}.  For a prescribed number of recovered packets $r$, the
degree distribution $D(r)$ returns the degree of the next output packet.

The scenario studied in \cite{GC} is a random encounter scenario, in
which each node meets independently at random one neighbor at a
time. The optimum degree distribution depends on the dynamics of the
underlying network and in~\cite{Munaretto2007Persistence} the authors
show the deficiencies of such an algorithm in scenarios beyond the
specific model they were designed for.  Since \cite{GC}, some other
coding algorithms based on a pre-defined degree distribution were
proposed. As a comparison algorithm, we use the following instance.\\

\textbf{Adaptive Network Coding (ANC)} \cite{Resilient_Coding}:
\emph{When a transmission opportunity occurs, the node randomly
  combines a specific number of packets in its buffer, which ensures
  that the degree of the resulting output packet is as high as
  possible and less than or equal to $D(r)$ (as defined above).}

\subsection{Opportunistic Algorithm}

The previous algorithms do not make use of feedback information in
terms of the recovery state of neighboring nodes. However, if
available, this additional information can be be used to make more
efficient coding decisions and can bring significant improvements to
the overall performance.  The work presented
in~\cite{Online_broadcasting_with_NC} analyzes how it can be used and
proposes a number of heuristic algorithms designed for the scenario of
a source node broadcasting to $n$ receivers over independent erasure
channels, i.e. a one-to-all broadcast scenario. Among the algorithms
proposed in \cite{Online_broadcasting_with_NC}, we consider the so
called Opportunistic algorithm, because it is the only one that can be
used in conjunction with a simple decoder. The basic idea is that each
node uses the feedback received from its neighbors to compute a queue
of symbols that have not yet been received by at least one node. The
first symbol is chosen randomly and further symbols are added under
the condition that the packet remains immediately decodable by all
neighbors that were previously able to decode it (in other words, the
number of nodes that can decode the packet can only increase).

Before defining a different set of algorithms capable of exploiting
the knowledge of the recovery status of neighboring nodes, we must
introduce some basic notation. In the following, we will denote the
node performing the coding algorithm by $x$. The set of neighbors of
node $x$ is denoted by $N_x=\{1,\dots,m \}$ and
$B_x=\{s_1,\dots,s_n\}$ denotes the set of symbols in the buffer of
node $x$. $B_j$ represents the set of symbols that are in the buffer
of node $j$, with $j\in N_x$. The set of symbols that are in the
buffer of $x$ and that are not in the buffer of neighbor $j$ is
denoted by $\overline{B}_j$ and can be computed as
$\overline{B}_j=B_x \backslash (B_j\cap B_x)$.

The algorithms presented next start by choosing the set of symbols $C$ that will be combined in the output packet by means of an XOR operation. The set $C$ is constructed iteratively, such that in each step of the algorithm one symbol is added to the previously constructed set. $\overline{C}$ denotes the set of symbols that node $x$ has in its buffer and that are not in $C$ (it can be calculated as $\overline{C}=B_x \backslash C$).

Finally, we will denote by $R(C)$ the set of neighbors of node $x$ that are able to recover a new symbol from the XOR of all symbols in a given set $C$. This means that, if there are $y$ symbols in set $C$, the neighbor must have already recovered exactly $y-1$ of the symbols in $C$. Therefore, we have that $R(C)=\{j: |C|-|B_j\cap C|=1 \}$.

We are now ready to give a formal definition of the Opportunistic scheme, presented in \cite{Online_broadcasting_with_NC}.
\begin{small}
\begin{algorithm}[Opportunistic Algorithm \cite{Online_broadcasting_with_NC}]
\caption{Opportunistic algorithm \cite{Online_broadcasting_with_NC}}
\label{oppor}
$_{}$\\
$_{}\ \ \ \ \ \ \ \ C=\emptyset$\\
$_{}\ \ \ \ \ \ \ \ S=\{s \in B_x: |R(\{s \})|>0 \}$\\
$_{}\ \ \ \ \ \ \ \ \textrm{while } |S|>0$\\
$_{}\ \ \ \ \ \ \ \ \ \ \ \ \ \ \ \ \textrm{choose } s^*\in S$\\
$_{}\ \ \ \ \ \ \ \ \ \ \ \ \ \ \ \ \textrm{add } s^* \textrm{ to } C$\\
$_{}\ \ \ \ \ \ \ \ \ \ \ \ \ \ \ \ S=\left(\bigcap \limits_{j\in R(C)} (B_j\cap B_x) \right) \cap \overline{C}$\\
$_{}\ \ \ \ \ \ \ \ \textrm{end}$\\
$_{}\ \ \ \ \ \ \ \ p=\bigoplus \limits_{s\in C} s$\\
$_{}\ \ \ \ \ \ \ \  \textrm{transmit $p$.}$
\end{algorithm}
\end{small}
As shown in Algorithm 1, the set $C$ is initialized to the empty
set. Throughout the algorithm, set $S$ represents the set of symbols
that can be added to the current configuration of set $C$. Initially,
all the symbols that are in the buffer of node $x$ and that are not in
the buffer of at least one neighbor are deemed to be possible
candidates. Thus, because $R(\{s\})$ is the set of neighbors that has
not recovered symbol $s\in B_x$, $S$ is initialized according to
$S=\{s \in B_x: |R(\{s \})|>0 \}$.

Next, we focus on the loop in the algorithm, which will only continue
while $S$ is non-empty. As long as there are symbols in $S$, the
algorithm chooses one of them randomly and adds it to $C$. After this
step, it is necessary to update $S$, from which the algorithm chooses
new symbols to add to $C$.

The new set $S$ is defined as the set of symbols that satisfy the
following conditions: (1) they are present in the buffers of the
neighbors in $R(C)$ (i.e. those neighbors that are able to recover a
new symbol from the current set $C$), (2) they are stored in the
buffer of node $x$; and (3) they were not chosen up to this
step. Thus, the new set $S$ can be determined according
to $$S=\left(\bigcap \limits_{j\in R(C)} (B_j\cap B_x) \right) \cap
\overline{C}.$$

When the loop finishes, all the symbols in set $C$ are XORed together
and the corresponding packet is sent to the neighbors of node $x$.

\begin{figure}[t]
  \includegraphics[width=0.45\textwidth]{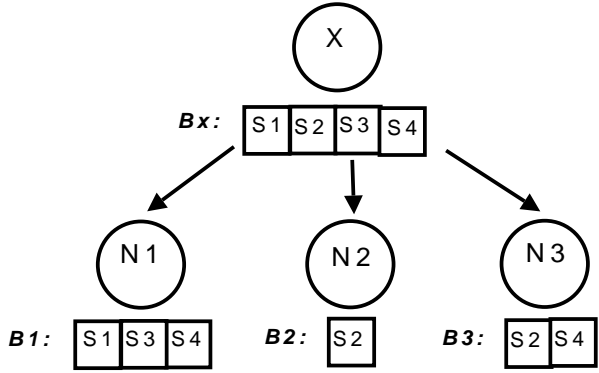}
  \caption{An example of a network, in which node ${\bf X}$ has to make coding decisions based on the buffer state of ${\bf 3}$ neighbors: ${\bf N_1}$, ${\bf N_2}$ and ${\bf N_3}$. For each neighbor, the figure depicts only that part of the buffer which contains symbols that are also stored in the buffer of node ${\bf X}$. }
  \label{fig:example}
  \vspace{-1.5em}
\end{figure}

The ideas behind the algorithm are best described by the example shown
in Fig.~\ref{fig:example}. Node $X$ (the node performing the coding
algorithm) has in its buffer symbols $s_1$, $s_2$, $s_3$ and
$s_4$. From Fig.~\ref{fig:example} it is clear that some of these
symbols can also be found in the buffers of the three neighbors of
$X$. It follows that there are only two possible output packets which
are optimal in the sense that they maximize the number of neighbors
able to recover a new symbol immediately upon reception: $p=s_1\oplus
s_2$ or $p=s_2\oplus s_3$. In both these cases, all neighbors are able
to recover a new symbol from the received packet, since they have one
and only one of the symbols therein. Moreover, no other combination of
symbols can provide a packet that immediately provides a new symbol to
every neighbor.

We now analyze the behavior of the Opportunistic algorithm. The
initial set $S$ (from which we can choose a symbol to be mixed in the
packet) is $S=\{s_1,s_2,s_3,s_4 \}$, due to the fact that none of the
neighbors has recovered all the symbols in the buffer of node
$X$. Thus, in the first iteration, each symbol in $S$ can be chosen
with probability $1/4$. Suppose that the algorithm chooses $s^*=s_1$
(again with probability $1/4$). Then, we have that $C=\{s_1\}$ and
$R(C)=\{N_2,N_3\}$. Recall that $R(C)$ is the set of neighbors that
have recovered all but one of the symbols in $C$ (in this case, it is
the set of neighbors who have not yet recovered symbol $s_1$). Since
$S$ is the set of symbols that (a) all the neighbors in $R(C)$ have
already recovered and (b) have not yet been chosen, we have that
$S=\{s_2\}$. In the second iteration, since $S=\{s_2\}$, the algorithm
chooses $s^*=s_2$ and sets $C=\{s_1, s_2\}$. Thus, $R(C)$ is
equivalent to the entire set of neighbors and, since there are no more
symbols recovered by the ensemble of neighboring nodes, we have that
$S=\emptyset$. Hence, the algorithm stops and outputs the packet
$p=s_1\oplus s_2$, which can be classified as an ideal packet.

We have seen that the algorithm outputs an ideal packet if the first
chosen symbol corresponds to $s_1$ (and that this happens with
probability $1/4$). Analogously, if the algorithm chooses $s_3$ first,
then $C=\{s_3\}$ and $R(C)=\{N_2,N_3\}$, yielding $S=\{s_2\}$. Hence,
in the next step the algorithm chooses symbol $s_2$ which will lead to
$S=\emptyset$. It follows that if the algorithm starts by choosing
symbol $s_3$, then we get the ideal output packet $p=s_2\oplus s_3$.

Suppose now that the algorithm starts by choosing symbol $s_2$. In
this case, we have that $C=\{s_2\}$ and $R(C)=\{N_1\}$. Since $S$ is
the set of unselected symbols recovered by the ensemble of neighbors
in $R(C)$, we have that $S=\{s_1,s_3,s_4\}$. Hence, in the second
iteration, the algorithm has a probability $1/3$ of choosing each of
the symbols in $S$. If the algorithm chooses symbol $s_1$
(respectively, symbol $s_3$), based on the same arguments as in the
previous cases, we deduce that the output packet will be $p=s_1\oplus
s_2$ (respectively, $p=s_2\oplus s_3$), which is an ideal packet. In
case the algorithm chooses $s_4$, the output will not be an ideal
packet.  Thus, the probability that the algorithm outputs an ideal
packet is given $$
\frac{1}{4}+\frac{1}{4}+\frac{1}{4}\left(\frac{1}{3}+\frac{1}{3}
\right)=\frac{2}{3}. $$

It is worth noting that in this algorithm, the sole criterion for the
choice of symbols to be mixed in the output packed is to ensure that a
node which is able to recover a new symbol from the current set $C$
(constructed up to the given iteration), will continue to be able to
recover a new symbol from the instances of $C$ that are constructed
after that iteration. In other words, after the choice of the first
symbol (which is performed randomly), the algorithm simply ensures
that the number of neighbors that are able to recover a new symbol
does not decrease with the next decisions.

\section{Optimized Coding Algorithms}
\label{sec:algorithms}

In the following, we present two algorithms for the encoding process, both based on the knowledge of the recovery status of the neighboring nodes. In order to increase the speed of information dissemination, our algorithms make coding decisions that by design allow the neighboring nodes to recover another information unit immediately upon reception of a new coded packet.

\subsection{Greedy algorithm}

The first algorithm gives priority to the symbols that are rarest within the neighborhood. The key is to find the combination of original symbols that maximizes the number of neighbors that are able to decode a new information unit.

\begin{small}
\begin{algorithm}[Greedy algorithm]
\caption{Greedy algorithm}
$_{}$\\
$_{}\ \ \ \ \ \ \ \ C=\emptyset$\\
$_{}\ \ \ \ \ \ \ \ \textrm{choose } s^* =\arg \max \limits_{s\in B_x} |R(\{s \})|$\\
$_{}\ \ \ \ \ \ \ \ q=0$\\
$_{}\ \ \ \ \ \ \ \ \textrm{while } |R(C\cup\{s^*\})|\geq q$\\
$_{}\ \ \ \ \ \ \ \ \ \ \ \ \ \ \ \ q=|R(C\cup\{s^*\})|$\\
$_{}\ \ \ \ \ \ \ \ \ \ \ \ \ \ \ \ \textrm{add } s^* \textrm{ to } C$\\
$_{}\ \ \ \ \ \ \ \ \ \ \ \ \ \ \ \ \textrm{choose } s^*=\arg \max \limits_{s\in B_x \backslash C} |R(C\cup\{s\})|$\\
$_{}\ \ \ \ \ \ \ \ \textrm{end}$\\
$_{}\ \ \ \ \ \ \ \ p=\bigoplus \limits_{s\in C} s$\\
$_{}\ \ \ \ \ \ \ \  \textrm{transmit $p$.}$
\end{algorithm}
\end{small}

As shown in Algorithm 2, the choice of the first symbol is very different from the Opportunistic algorithm ({\it Algorithm
\ref{oppor}}). Instead of a random choice, the Greedy algorithm selects the symbol that maximizes the number of nodes that are able to decode a new symbol if a packet of degree one is sent. This corresponds to maximizing $|R(\{ s\})|$. If there are
multiple symbols that satisfy this condition, the algorithm chooses one of
them randomly. As we will see later on, a proper choice of the first
symbol is crucial for a good performance. In fact, we will show that,
if the nodes send packets of degree one (i.e. plain symbols) and use the selection criteria of our protocols, the resulting performance is already quite close to the performance of the Opportunistic algorithm.

Taking a closer look at the loop of this algorithm, we realize that after choosing the
first symbol, the algorithm proceeds by selecting among the symbols yet to be
chosen the element that maximizes the number of neighbors 
able to decode a new symbol from the packet, which is obtained by XORing this
new symbol with all the symbols selected so far. This can be written as 
$|R(C\cup \{s \})|$. After choosing this candidate symbol (a symbol $s^*=\arg \max \limits_{s\in B_x \backslash C}
|R(C\cup\{s\})|$), the algorithm will check if there is a gain in
adding this candidate symbol to the set of symbols to be mixed in the output packet. Notice that, for the algorithm to continue, we do not require that neighbors that could previously recover a new symbol will continue to be able to do so; the algorithm continues as long as the number of neighbors able to recover a new symbol does not decrease from one
step to the next one.

Denote by $p^*$ the packet obtained by XORing all the symbols chosen so far (i.e., all the symbols in the current set $C$) and denote by $s^*$ the candidate symbol. If the number of neighbors that are able to decode a new symbol from $p^*\oplus s^*$ (which is
represented by $|R(C\cup \{s^* \})|$) is less than the number of
neighbors that are able to decode a new symbol from $p^*$ (which is
represented by $q$), i.e. if $|R(C\cup \{s^* \})|\geq q$, the
algorithm stops and produces a packet that combines all of the symbols selected thus far.

Going back to the scenario illustrated in Fig.~\ref{fig:example}, we see that the algorithm starts by choosing the symbol $s^*$ that maximizes the size of $R(\{s\})$ over all $s$ in the buffer of node $X$, i.e. that maximizes the number of neighbors that do not have the symbol $s^*$. Clearly, $s^*$ is the {\it rarest symbol in the neighborhood}. It follows that the algorithm ends up choosing $s_1$ or $s_3$, since each of them is present in the buffer of only one of the neighbors. If the algorithm chooses symbol $s_1$, we have that $|R(\{s_1\})|=2$. In the first iteration, the algorithm sets $q=2$ and $C=\{s_1\}$. Next, the algorithm selects the symbol $s^*$ as the one that maximizes the size of $R(C\cup \{s\})$ over all $s\neq s_1$. More specifically, it will choose the symbol that maximizes the number of neighbors that are able to recover a new symbol from the packet obtained when XORing this candidate symbol with all the symbols in $C$. In this case, since $C=\{s_1\}$ and all $3$ neighbors can recover a new symbol from $s_1\oplus s_2$, this candidate symbol is $s^*=s_2$. Now, the algorithm checks if the number of neighbors that can recover a new symbol increases when compared to the previous step. In this case, since $q=2$ neighbors recovered a new symbol and adding the candidate symbol increases this number to $3$ (i.e. $|R(C\cup \{s_2\})|\geq 2$), the algorithm continues by updating $q$ to $q=3$ and adding the candidate symbol to the packet: $C=\{s_1,s_2\}$. Now, the algorithm chooses the next candidate symbol using the same rule, i.e. to maximize the number of neighbors that are able to recover a new symbol. In this case, this symbol can be $s_3$ or $s_4$. In either case, we have that only one neighbor will be able to recover a new symbol if the candidate symbol is added, thus we will have $|R(C\cup \{s^*\})|=1$. In the subsequent step, since $|R(C\cup \{s^*\})|<3=q$, the algorithm stops and outputs the packet $p=s_1\oplus s_2$, which is an ideal packet.

Notice that in the first choice we had two options: $s_1$ and $s_3$. We saw that if $s_1$ is chosen, the algorithm outputs the ideal packet $p=s_1\oplus s_2$. Using analogous arguments, it is easy to see that if $s_3$ is chosen in the first step, the algorithm outputs the packet $p=s_2\oplus s_3$, which is also an ideal packet. Thus, we have that in this example, with probability $1$, the Greedy algorithm outputs an ideal packet.

Similarly to the Opportunistic algorithm, the Greedy algorithm evolves in each iteration by selecting a symbol to be added to the set of symbols that will form the output packet.  After the choice of the first symbol, the algorithm ensures that the number of neighbors that are able to decode new symbols does not decrease with the next decisions. Beyond the choice of the first symbol (which has a significant impact on the performance as we will see
latter on), the selection procedure targets the symbol that will maximize the number of neighbors that are able to decode, whereas the Opportunistic algorithm make this selection in a purely random fashion.

\subsection{Equalizing algorithm}

The Greedy algorithm presented in the previous section is prone to lead to an uneven distribution of information. In the worst case, some nodes that are not well connected to the rest of the network might receive mostly packets they cannot decode, since they lack some of the information units that all the other nodes already have. These nodes would be served by the greedy algorithm only after all other nodes have decoded all of the information, leading to a high worst case delay. The way to prevent this from happening is to equalize the recovery level among the neighbors instead of maximizing it in a greedy fashion.

The so called Equalizing algorithm pursues mainly the goal of giving new decodable information to the neighbors that have recovered the fewest information units, thus increasing the minimum number of recovered packets per node.

\begin{small}
\begin{algorithm}[Equalizing algorithm]
\caption{Equalizing algorithm}
$_{}$\\
$_{}\ \ \ \ \ \ \ \ C=\emptyset$\\
$_{}\ \ \ \ \ \ \ \ B=B_x$\\
$_{}\ \ \ \ \ \ \ \ R^*(C)=\{j: C \subseteq B_j \}$\\
$_{}\ \ \ \ \ \ \ \ \textrm{while } |B|>0 \textrm{ and } |R^*(C)|>0$\\
$_{}\ \ \ \ \ \ \ \ \ \ \ \ \ \ \ \ \textrm{choose } J=\arg \min \limits_{j\in R^*(C)} |B_j|$\\
$_{}\ \ \ \ \ \ \ \ \ \ \ \ \ \ \ \ S=B\cap \overline{B}_J$\\
$_{}\ \ \ \ \ \ \ \ \ \ \ \ \ \ \ \ \textrm{choose } s^*=\arg \max \limits_{s\in S} |R  (C\cup \{s \})|$\\
$_{}\ \ \ \ \ \ \ \ \ \ \ \ \ \ \ \ \textrm{add } s^* \textrm{ to } C$\\
$_{}\ \ \ \ \ \ \ \ \ \ \ \ \ \ \ \ B=B \cap B_J$\\
$_{}\ \ \ \ \ \ \ \ \textrm{end}$\\
$_{}\ \ \ \ \ \ \ \ p=\bigoplus \limits_{s\in C} s$\\
$_{}\ \ \ \ \ \ \ \  \textrm{transmit $p$.}$
\end{algorithm}
\end{small}

IN Algorithm 3, $R^*(C)=\{j: C \subseteq B_j \}$ represents the set of
neighbors that have all the symbols in $C$. In each step, the
algorithm chooses the neighbor that has the least recovered packets
among those not yet considered.  Then, the algorithm selects one of
the symbols that this particular neighbor has not yet recovered (and
that all the previously chosen neighbors did recover, thus ensuring
that the previously chosen neighbors can still decode the
packet). This symbol is added to the packet to be sent.

The algorithm needs to keep track of the symbols that neighbors chosen so far have already recovered. This is captured by set $B$. One condition to stop the loop of the algorithm is precisely the existence of symbols in $B$. If there are no symbols $B$, i.e. if there is no symbol that has been recovered by all the nodes chosen up
to a certain iteration, no symbol can be added to the packet to be sent without rendering at least one of the neighbors unable to decode. The other condition for the loop to stop is $|R^*(C)|>0$, which means that the loop only continues if there are still neighbors that have recovered all the symbols in the output packet constructed so far. If there are no neighbors in this condition, no further nodes will be able to recover a new symbol irrespective of which symbol is added to the packet.

In each iteration, the algorithm starts by inspecting all nodes that have recovered all the symbols in the packet constructed so far (i.e. neighbors in $R^*(C)$ which implies that no neighbor can be chosen twice) and finding the one that recovered the least number of symbols. More specifically, we choose the neighbor $J$ that satisfies $J=\arg \min \limits_{j\in R^*(C)} |B_j|$. After making this selection, the algorithm calculates the set of symbols that can still be added to the packet. These symbols must have been recovered by all the previously chosen neighbors(i.e., symbols in $R$) and cannot have been recovered by the neighbor that was chosen in the current iteration (i.e., symbols not in $B_J$). Thus, the set of candidates is defined by $S=R\cap \overline{B}_J$.

Next, from this set of candidate symbols, the algorithm selects the one that maximizes the number of neighbors that are able to decode a new symbol, assuming that the output packet results from the XOR of all symbols in $C$.
After this choice, the algorithm adds the symbol to the set $C$ and updates the set $R$. The new set $R$ will be the set of symbols shared by all the neighbors that were chosen before the current iteration (namely $R$) and possessed by the new chosen neighbor ($B_J$), i.e. $R=R\cap B_J$. When the loop is completed, the algorithm computes the packet to be sent by XORing all the symbols in $C$.

Once again, we will use the scenario in Fig.~\ref{fig:example} to clarify the main steps of the algorithm. The algorithm starts by setting $C=\emptyset$, $B=\{s_1,s_2,s_3,s_4\}$ and $R^*(C)=\{N_1,N_2,N_3\}$ (recall that $R^*(C)$ represents the set of neighbors that have already recovered {\it all} of the symbols in $C$). In the first iteration, the algorithm starts by checking which neighbor has the smallest buffer, i.e. the one with the smallest number of recovered symbols. In this case, the chosen neighbor is $N_2$, since it only recovered symbol $s_2$. Then, the algorithm computes the set of symbols that this node does not have in its buffer: $s=\{s_1,s_3,s_4\}$. The goal is to provide a new symbol to this particular neighbor. Hence, the first chosen symbol is a symbol from $S$ and, since we also want to provide (if possible) new symbols to other neighbors, the algorithm chooses the symbol that is more rare within the neighborhood, among all the symbols in $S$. In this case, we have two options: $s_1$ or $s_3$. Suppose that the algorithm chooses $s^*=s_1$. We have that $C=\{s_1\}$ and $B=\{s_2\}$, i.e. $B$ is the set of symbols that node $N_2$ has already recovered. It is necessary to keep track of the symbols that all the neighbors chosen by the algorithm have already recovered to ensure that the neighbors with the smaller number of recovered symbols will be able to recover a new symbol from the resulting output packet.

Next, in the second iteration, the algorithm chooses the neighbor that
has the smallest number of recovered symbols among all the neighbors
that have all the symbols in the current instance of set $C$,
i.e. among the neighbors in $R^*(C)$. In this case, $R^*(C)=\{N_1\}$
and thus the chosen neighbor is $N_1$. Now, the set of symbols that
can be added to $C$ is the set of all symbols that all the previously
chosen neighbors have in their buffers, $B$, and that the neighbor
chosen in the current iteration does not have in its buffer,
$\overline{B}_1$. Thus, in this case, we have that $S=\{s_2\}$ and,
hence, $C=\{s_1,s_2\}$. Notice that there are no further symbols that
have been recovered by all the chosen neighbors, i.e. $B=\{s_2\} \cap
\{s_1,s_3,s_4\}=\emptyset$. Thus, the algorithm cannot continue and
consequently outputs the packet $p=s_1\oplus s_2$, which is an ideal
packet.

In the first iteration, we could have chosen symbol $s_3$ instead of
$s_1$. Using similar arguments, it is easy to see that, if $s_3$ is
chosen, the algorithm outputs the packet $p=s_2\oplus s_3$, which is
also an ideal packet. Therefore, we again have that with probability
$1$ the Equalizing algorithm outputs an ideal packet in our example.

\section{Simulation Results}
\label{sec:results}

In this section we present and discuss the performance of the
aforementioned coding algorithms in various scenarios. The main
part of the analysis assumes the simple decoding
algorithm.

We discuss the enhancement of performance provided by the use of a
full decoding scheme at the end of this section.  The following
performance metrics are of interest for the analysis of the
algorithms. The recovery rate of a node is the number of original
packets recovered by the node as a function of the total number of
packets received by the node. It measures the speed of the
dissemination process and thus the overall efficiency of the
protocol. This metric is crucial in communication networks, especially
for delay-sensitive applications such as real-time media streaming,
where the disseminated segments must be recovered by the receivers
within strict time intervals.  We average the recovery rates over all
nodes and repeated the simulations several times so as to get tight
confidence intervals. For the recovery rate shown in the next plots,
the confidence intervals are all within $\pm 2 \%$. These intervals
are omitted from the figures for the sake of readability.

The codeword degree is measured as the number of original symbols
which are combined to form a (coded) packet and we calculate
the {\it average codeword degree} over all nodes.  For the analysis of
the single-hop scenario we also consider {\it packet delay} as defined
in Definition 2 of \cite{Online_broadcasting_with_NC}. The delay that
a receiver experiences is the total number of received packets that do
not allow immediate recovery of a new original symbol. Again we
consider this delay averaged over all nodes.  Finally, the {\it
information potential} is the number of original symbols that are
available at a given neighboring node, but not at the node itself. This
metric is useful for analyzing the overlap between the information
recovered by a target node and by its neighbors. It provides a measure
of the efficiency of the coding process and gives an insight into the
degree of freedom available for making coding decisions.

The simulation results have been obtained using a custom C++
simulator. It provides an ideal (collision-free) MAC layer, with a
sequential or random scheduling of packets. All transmissions use
physical layer broadcast.

\subsection{Single-hop Scenario}

In this setting, a single source node broadcasts $100$ original
symbols to its $100$ neighbors over independent erasure channels and,
as in \cite{Online_broadcasting_with_NC}, perfect feedback is
available from the receivers to the source. For all the algorithms in
our analysis, i.e. Greedy, Equalizing, Opportunistic and ANC, the
source node starts by sending out all the original symbols in uncoded
form. It is only after this initial stage that the source node sends
encodings of the original symbols as described in Sections
\ref{sec:related} and \ref{sec:algorithms}. The erasure probability is
set to $0.5$.

\begin{figure}[t]
  \includegraphics[width=0.51\textwidth]{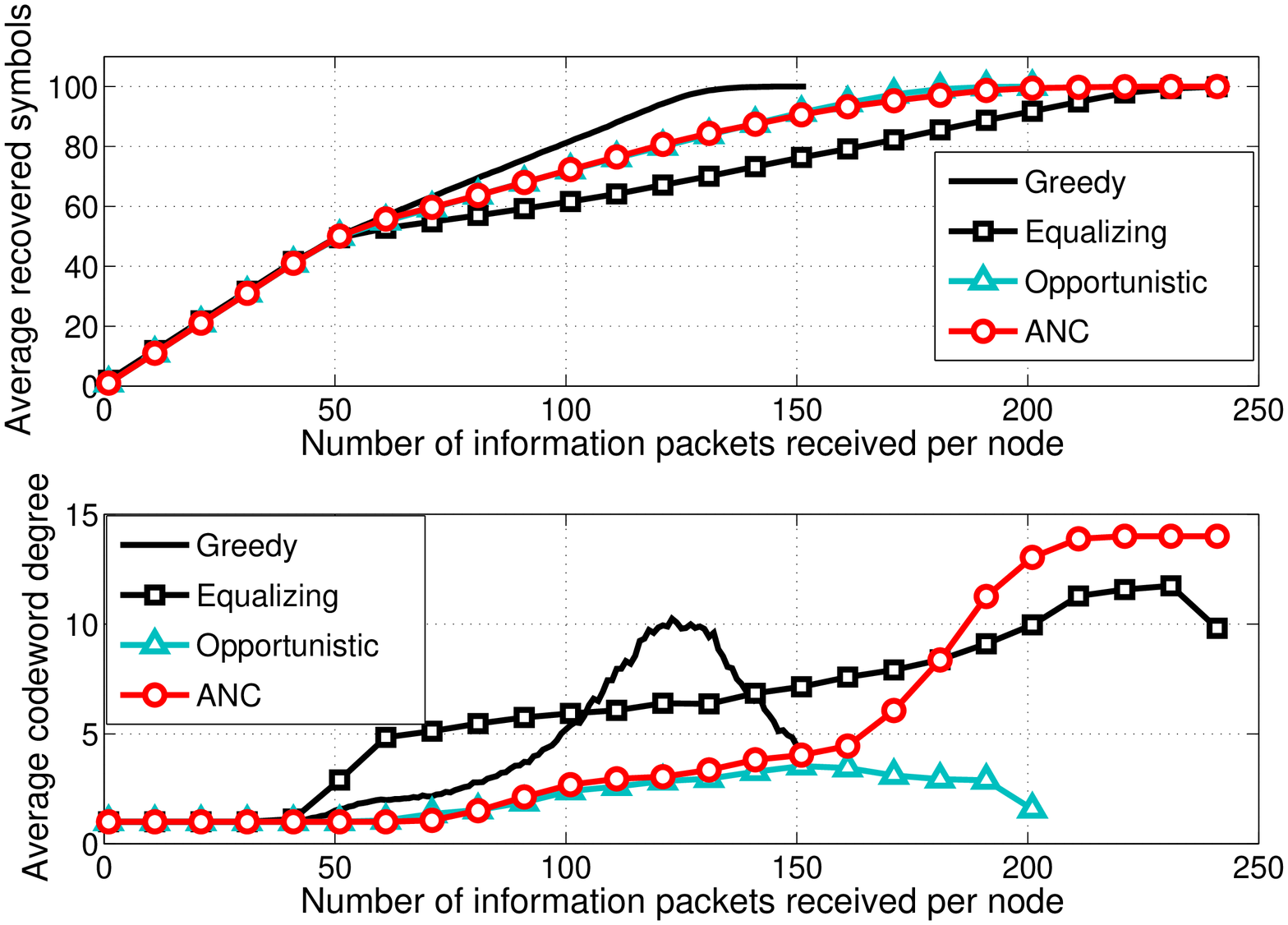}
  \vspace{-1.5em}
  \caption{Recovery rate (top) and average codeword degree (bottom) for the single-hop scenario.
  }
  \label{1hop}
  \vspace{-1.5em}
\end{figure}

In Fig.~\ref{1hop}, the Greedy algorithm shows the best performance in
the single-hop scenario. With this algorithm, all nodes achieve the
full recovery of the $100$ original symbols within $150$ information
packets received. From the first 100 uncoded packets the receivers
miss around 50 original packets and they differ from node to node due
to the random erasure pattern.  This allows the Greedy algorithm to
increase the degree of the coded packets compared to the Opportunistic
and ANC algorithms between 50 received packets and 120 received
packets. When the process approaches the full recovery state, the
number of nodes still missing some packets decreases and low
degree codewords are sufficient to serve these nodes. Given the diversity of
missing information among all the receivers and the large amount of
information potential of the neighbors (due to the erasure pattern),
the degree of freedom for making coding decisions lets the source node
perform the coding that best allows a large number of receivers to
immediately recover a new symbol from the sent packet.

The Equalizing algorithm has a considerably worse recovery rate than
the other algorithms. The reason is visible in Fig.~\ref{1hop},
bottom, where Equalizing starts using higher and higher codeword
degrees quite early on. Since it is designed to provide an immediately
decodable packet to the neighbor(s) which recovered the least number
of original symbols, many other neighbors are not able to decode the packet
--- their composition of recovered packets differs from those poor
nodes. Focusing only on the poor nodes results in a packet that despite
its high degree is useful only for few receivers.

Surprisingly, the performance of ANC and of the Opportunistic
algorithm is almost the same for most of the simulation. The
Opportunistic algorithm, which allows the source node to use the
neighborhood status information to make the coding decisions, performs
just slightly better than ANC.  Due to the large information potential
of the neighbors and to the huge diversity of missing information
among the receivers, the choice of which symbols to encode is not
crucial, as long as the number of symbols that are combined is the
same. Only at the very end, the information obtained from the
receivers by the Opportunistic algorithm allows the source node to
send the last few missing symbols without wasting time sending packets
that carry encodings of randomly picked symbols.

Focusing the analysis on the {\it packet delay}, we note that the
Greedy algorithm achieves the lowest packets delay of only $30$ not
useful packets (averaged over all receivers), followed by the
Opportunistic algorithm with $75$, ANC with $90$ and Equalizing with
$100$, as could be expected from the previous analysis.

\subsection{Multi-hop Scenario}

In the multi-hop scenarios, each of the $100$ nodes generates one
original symbol that is intended to be delivered to every other node
in the network.

\subsubsection{Static grid}

\begin{figure}[t]
\includegraphics[width=0.51\textwidth]{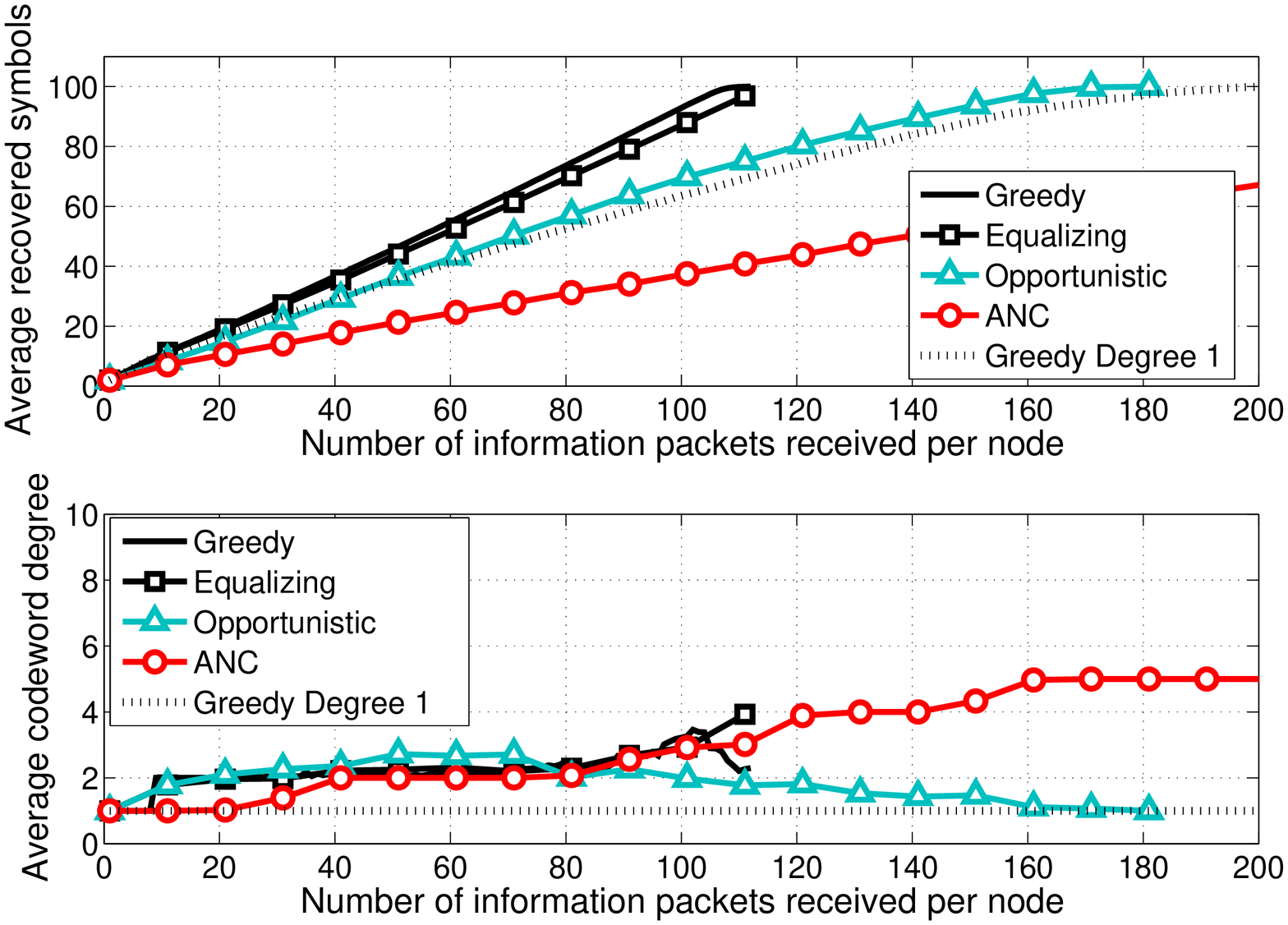}
     \vspace{-1.5em}
  \caption{Recovery rate (top) and average codeword degree (bottom) for 100 nodes on a static grid.
  }
  \label{grid_wrap}
   \vspace{-1.5em}
\end{figure}

In this setting, nodes are placed on a static grid (that wraps around)
and each node has eight neighbors to communicate with. In
Fig.~\ref{grid_wrap}, top, the algorithm with the best performance is
the Greedy one (as in the previous scenario), but now the difference to the
performance of the Equalizing algorithm is much smaller. From the
analysis of their respective average codeword degrees, in
Fig.~\ref{grid_wrap}, bottom, we see that the coding degree of our two
algorithms is very similar, except for the very end where the Equalizing
algorithm takes longer than the Greedy algorithm to increase 
the coding degree for recovery of the last missing symbols.

The high degree of correlation of the information recovered by the
neighbors, due to the wrap around and the symmetrical topology, and
the consequently minor diversity of information stored by the
neighbors compared to the single-hop setting, makes the use of packets
with high degree ineffective. Moreover, choosing which original
packets to combine has a huge impact on the performance of the
dissemination process. To visualize this, we also show the recovery
rates achieved by the algorithms corresponding to Greedy and
Equalizing when we limit the codeword degree to one, i.e., only an
uncoded packet is sent. In the top graph of Fig.~\ref{grid_wrap}, we
plot only Greedy with codeword degree one, but both of the algorithms
perform the same. Even in this limited case, the recovery rates of our
algorithms are very close to the recovery rate of the Opportunistic
algorithm \emph{with} coding. The few degrees of freedom for making
coding decisions, typical of this setting, limit the performance of
the Opportunistic algorithm, where the first symbol is chosen
randomly.

Finally, the impact of using neighborhood recovery status in the
coding decisions is obvious when we compare the recovery rate of the
ANC algorithm with the recovery rates of the other algorithms.  For
instance, the total number of received packets necessary to achieve
full recovery is, in the case of ANC, several times larger than in the
case of Greedy, while the average codeword degree is quite similar for
most of the values of received packets.

\subsubsection{Static random and clustered networks}

\begin{figure}[t]
  \includegraphics[width=0.51\textwidth]{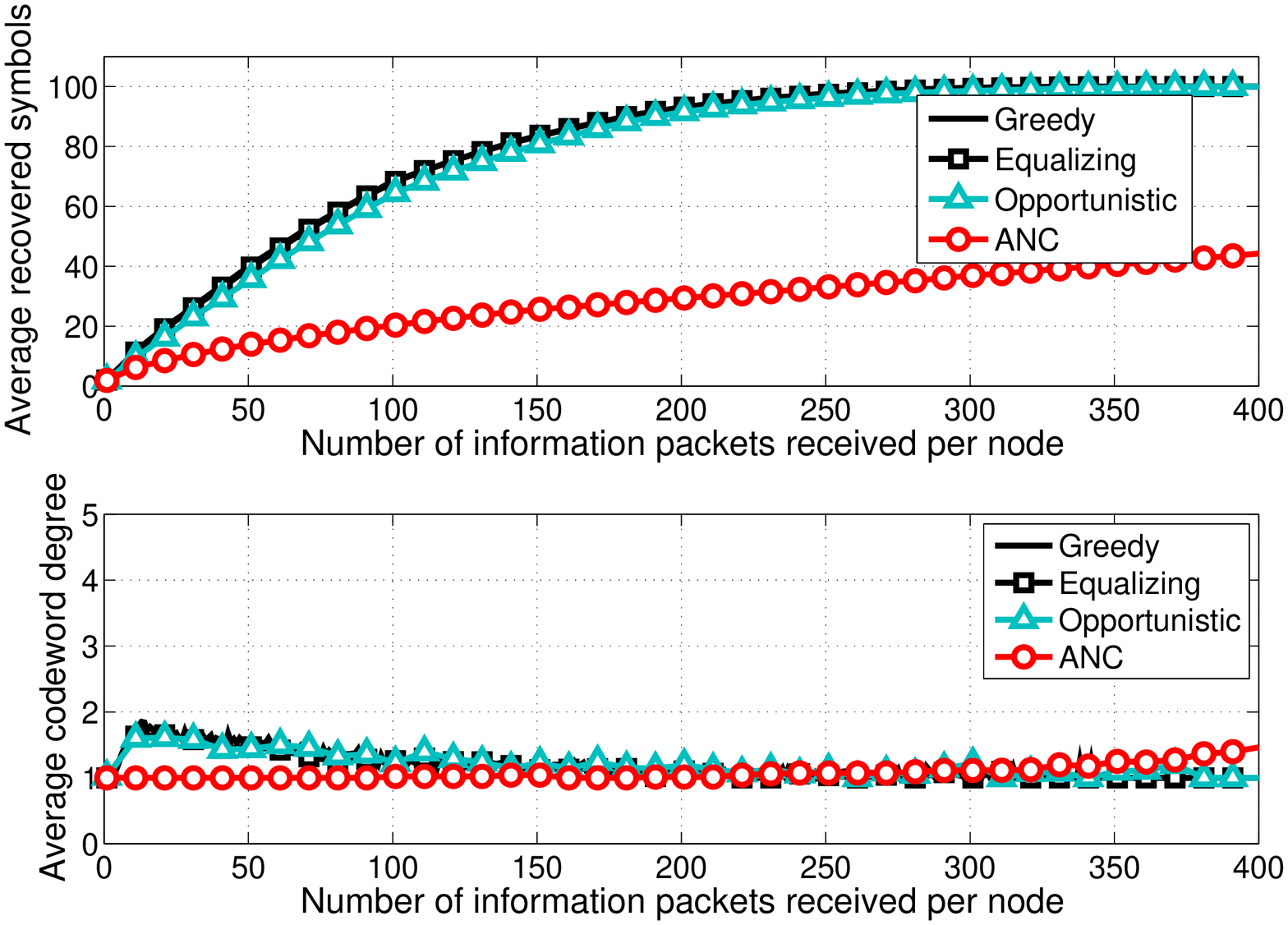}
   \vspace{-1.5em}
  \caption{Recovery rate (top) and average codeword degree (bottom) for the static random network, 100 nodes.
  }
  \label{random}
   \vspace{-1.5em}
\end{figure}

\begin{figure}[t]
\includegraphics[width=0.51\textwidth]{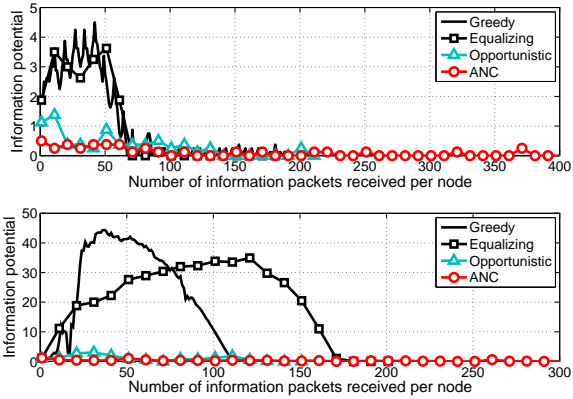}
   \vspace{-1.5em}
  \caption{Neighborhood information potential for static random network (top) and moderate mobility scenario (bottom).}
  \label{overlap}
   \vspace{-1.5em}
\end{figure}

In this section, we consider two different scenarios: static random
topologies with an average density of 8 nodes per communication range,
and clustered networks. These static networks are relatively sparse,
which means that the information potential of neighbors is smaller
than in the grid network and that the diversity of information stored
by the neighbors is lower. The high degree of correlation among the
original symbols recovered by the nodes explains why the Greedy,
Equalizing and Opportunistic algorithms perform similarly
(Fig.~\ref{random}). No degree of freedom for making specific coding
decisions is left to these algorithms, so that the differences are
small. Also in such settings, ANC cannot perform well given the
extremely low level of diversity of information among nodes.  In
Fig.~\ref{overlap}, top, we show the low information potential.  Only
Greedy and Equalizing increase the diversity of information among
neighbors at the beginning of the simulations, as expected from the
description of their coding mechanism in Section
\ref{sec:algorithms}. After ~$75$ packets received, all algorithms
experience the same neighborhood information potential (due to the
natural progressive lowering of the diversity of information over
time).  The differences of the performance in terms of recovery rate,
among Greedy, Equalizing and Opportunistic algorithms increase
slightly for larger node densities (not shown here). Such settings are
closer to the characteristics of a grid topology concerning the
degrees of freedom for making coding decisions.

The delay experienced by nodes using the Greedy and Equalizing
algorithms in static sparse random and clustered networks shows an
interesting result. For each setting in analysis, the average delay is
practically the same for the two algorithms, however the worst delay
(i.e. the delay experienced by the node with the highest delay) of the
Greedy algorithm is up to $5$\% higher than the worst delay of the
Equalizing algorithm.  Since the Equalizing algorithm will always try
to provide an immediately decodable packet to the neighbor with the
lowest number of recovered original symbols, this will obviously
improve worst case delay.

\subsubsection{Moderate mobility}

In this scenario, we consider nodes moving according to a random
waypoint mobility model with speeds uniformly distributed in the
interval $[2,4]\: m/s$. Again, the node density  allows on average for eight
neighbors per node.  We assume perfect information about the neighbor recovery status.

\begin{figure}[t]
  \includegraphics[width=0.51\textwidth]{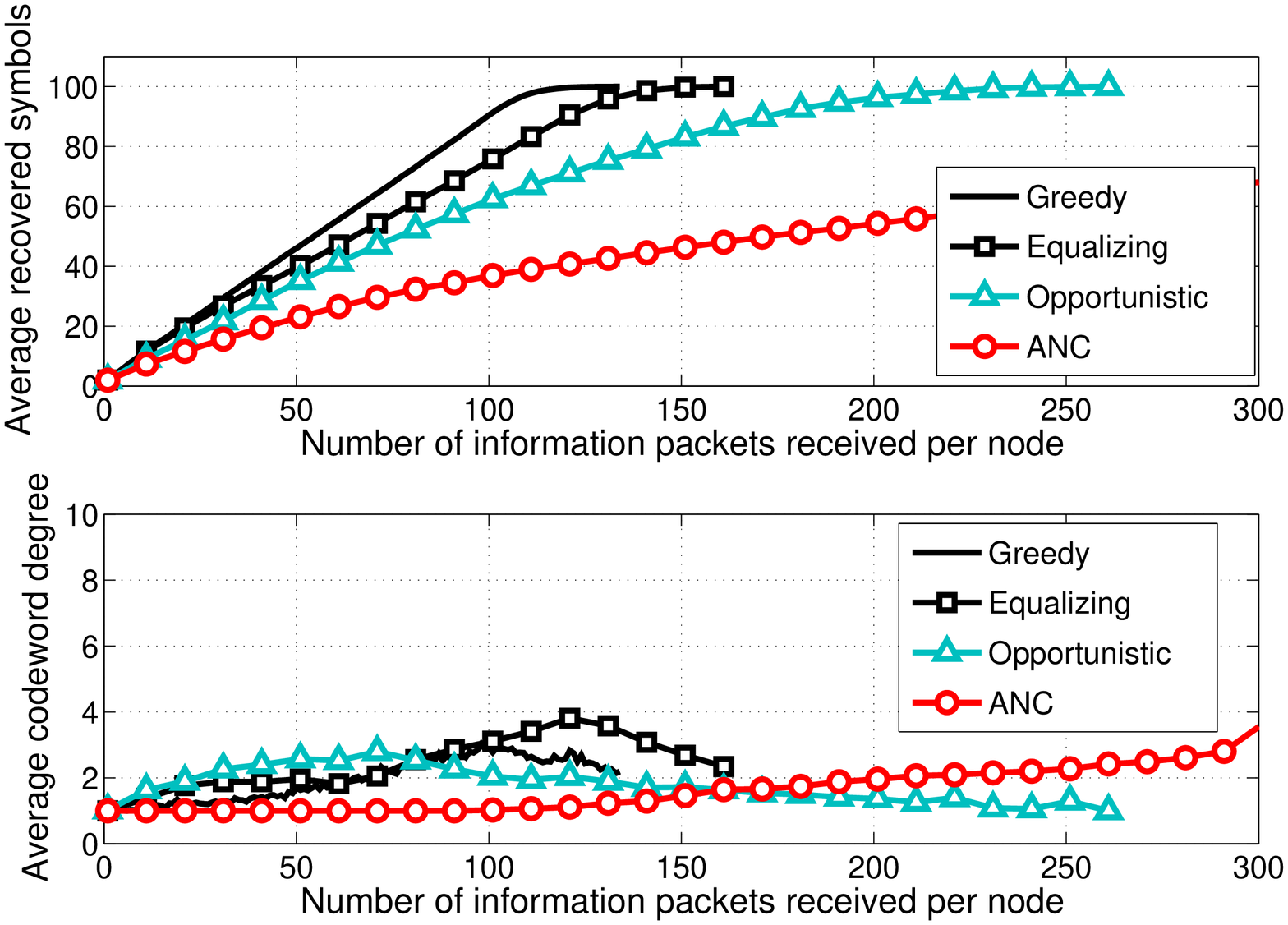}
   \vspace{-1.5em}
  \caption{Recovery rate (top) and average codeword degree (bottom) for the mobile scenario, 100 nodes.
  }
  \label{mobile}
   \vspace{-1.5em}
\end{figure}

In Fig.~\ref{mobile}, top, we notice that the performance of the
algorithms under consideration in terms of recovery rate is somewhat
similar to the one observed in the static grid setting (Fig.
\ref{grid_wrap}).  Due to the mobility of the nodes, the correlation
among the original symbols recovered by neighbors is much lower in the
case of moderate mobility than in the case of a static grid or static
random networks. An insight into the differences of information
potential of neighbors for two extreme cases is given in
Fig.~\ref{overlap}: on the top the random static network shows a lower
diversity of information among neighbors than in the mobile case,
bottom, where the information potential with our algorithms can
achieve very high values of up to $40-50$, whereas for the other
protocols the measured values are close to $0$. As seen in the
previous section, our algorithms allow nodes to maintain a high
diversity of information in the neighborhood and thus a high degree of
freedom for the coding decisions.

It is also important to notice that the coding degree of the
Equalizing algorithm is always higher than the coding degree of the
Greedy algorithm. This observation and the fact that the recovery rate
of the Greedy algorithm is higher than the recovery rate of the
Equalizing algorithm let us conclude that the Equalizing algorithm
builds packets with a too high codeword degree, rendering these
packets not immediately decodable. We will see later on in this paper
that, if we allow the use of a full decoding process, i.e., all the
packets in the buffer (decoded and undecoded) are considered for the
decoding algorithm, the recovery rate of the Equalizing algorithm can
actually surpass the recovery rate of the Greedy algorithm.

\subsection{Performance gains using a complete buffer decoding mechanism}

In this section, we investigate the benefits of full decoding, which
is more efficient (in the sense that it does not discard useful packets) 
but also more costly in terms of energy, memory
requirements, and processing. Up to now, we were considering a scheme
were only the immediately recovered original symbols were considered
for the simple decoding process. Here, all the packets received
(decoded and undecoded) are taken into consideration when performing
the decoding of the received packets.  In the following figures, we
omit the plot of the average codeword degree for the sake of
readability of the recovery rate. Also, the average codeword degree of
the algorithms using a full decoding scheme is almost the same as the
one with the simple decoder, except for a slight increase of the
average codeword degree.

\begin{figure}[t]
\includegraphics[width=0.51\textwidth]{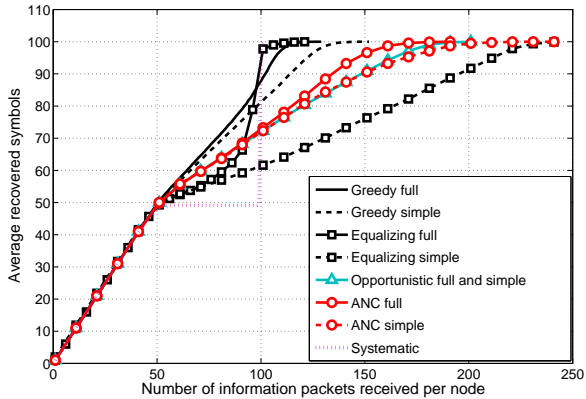}
   \vspace{-1.5em}
  \caption{Recovery rate for the single-hop scenario with simple and full decoding schemes.
  }
  \label{1hop_gauss}
   \vspace{-1.5em}
\end{figure}

As we already mentioned in the previous analysis, it is expected that
the recovery rate significantly increases with the full decoding
scheme, since the algorithms often produce packets that are not
immediately decodable for some neighbors but that are innovative. The
node is not able to recover a new original symbol from the received
information packet since it did not yet recover the required other
original symbols that form the coded information packet. By storing
these not immediately decodable but innovative packets in the buffers
and considering these packets in the decoding process, nodes can find
these packets helpful later on, when more and more (innovative or
immediately decodable) packets are received, increasing the recovery
rate of the algorithm. This benefit can be observed in
Fig.~\ref{1hop_gauss}, where the recovery rate of the algorithms using
a full decoding mechanism (including the Systematic Random Network
Coding algorithm) is plotted for the single-hop scenario.

Comparing the results obtained with this full decoding scheme to the
results obtained with a simple decoding scheme, we can observe a major
improvement of the recovery rate of the Equalizing algorithm.  With
the full decoding scheme, the Equalizing algorithm has a recovery rate
that slowly increases after the initial phase (where the nodes first
send the original symbols uncoded once) and, at around $80$ packets
received, shows a ``smooth'' step behavior that is typical of the
random network coding algorithms (e.g.  Systematic Random Network
Coding). The Equalizing algorithm reaches the full recovery state
before the Greedy algorithm. However, the recovery rate of the Greedy
algorithm is higher than the one of the Equalizing algorithm before
the step behavior takes place.

We also consider the packet delay when using a full decoding
mechanism. The delay presented by Greedy with this decoding scheme is
$10$ lower than the delay obtained when using the simple decoder, with
a total averaged delay of $20$. In \cite{Online_broadcasting_with_NC},
none of the algorithms proposed for the single-hop scenario is able to
achieve such a low delay (in the same conditions as used here). The
Equalizing algorithm experiences only an average delay of $40$
packets, ANC of $70$ and the Systematic Random Network Coding of $50$
as in \cite{Online_broadcasting_with_NC}.

After analyzing the performance enhancements achieved by using a full
decoding scheme in the single-hop scenario, we now discuss the results
obtained for the multi-hop scenario with moderate mobility.  We have
chosen this particular setting of the multi-hop scenario because the
performance of the algorithms in the other settings is similar to the
one presented in Fig.~\ref{grid_wrap}, although there are some
differences that should be pointed out. With a full decoding scheme,
the Greedy and Equalizing algorithms have quite similar
performance. Moreover, the enhancements achieved by the other
algorithms when using a full decoding scheme are negligible except for
the ANC algorithm, for which the performance is still far from the
performance achieve by our algorithms.

\begin{figure}[t]
\includegraphics[width=0.51\textwidth]{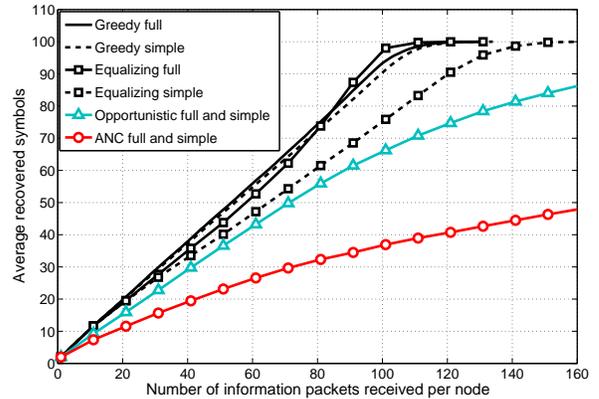}
   \vspace{-1.5em}
  \caption{Recovery rate for the multi-hop mobile network, with simple and full decoding schemes.
  }
  \label{mobile_gauss}
   \vspace{-1.5em}
\end{figure}

In Fig.~\ref{mobile_gauss}, we can again see that in the moderate
mobility scenario and with a full decoding scheme, there is a major
improvement of performance of the Equalizing algorithm. The recovery
rate of the Equalizing algorithm comes very close to the recovery rate
of the Greedy algorithm until around $80$ packets received (similar to
the behavior in the single-hop scenario). After this value, the
Equalizing algorithm is faster in recovering new original packets,
reaching the full recovery state $10$ packets before the Greedy
algorithm.  It is also interesting to notice that there is no
significant difference in terms of recovery rate between the
two decoding mechanisms in combination with the Greedy
algorithm. This behavior was
expected, since the Greedy algorithm was designed for immediate
decoding. Few not immediately decodable packets mean that a complete
buffer decoding mechanism can just slightly outperform a simple
decoder.

\section{Conclusions}
\label{sec:conclusions}

In this paper we proposed two coding algorithms which exploit 
feedback information on the recovery status of neighboring nodes. Through the
analysis of a wide range of settings in our simulations, we show that
the Greedy algorithm consistently outperforms all other algorithms in
terms of number of immediately decodable packets, which is fundamental
for delay-sensitive applications in wireless networks such as
real-time media streaming. Moreover, satisfactory results of the
Greedy algorithm are already obtained using just a simple decoder,
whereas for the Equalizing algorithm the use of Gaussian elimination
improves the performance significantly. However, using the Equalizing
algorithm is beneficial in some inhomogeneous (clustered) topologies,
where the worst case delay is lower than that of the Greedy algorithm
at a similar recovery rate.

The algorithms proposed in this paper focus on immediate decodability,
and hence take the decoding delay as the sole optimization criterion. As a next step, we
intend to explore the design tradeoff between delay and throughput in
more detail. The perceived quality of a video transmissions is largely dependent on
the right balance between the two. An algorithm which imposes slightly
less stringent delay requirements and allows decoding after reception
of a fixed number of packets (as opposed to just one packet), will provide higher
throughput, which may improve the overall perceived quality. We
further assumed instantaneous feedback in the evaluation of the
algorithms. The analysis of the impact of imperfect and delayed
feedback on the performance of network coding algorithms is left as an important 
item for future work.

\bibliographystyle{IEEEtran}
\bibliography{article}

\end{document}